\documentclass[10pt,a4paper,twocolumn]{article}
\usepackage[top=30mm, bottom=30mm, left=15mm, right=15mm]{geometry}
\usepackage{amssymb,amsmath,latexsym}
\usepackage[font={small,it}]{caption}
\usepackage[T1]{fontenc}
\usepackage[utf8]{inputenc}
\usepackage{amsmath}
\usepackage{amssymb}
\usepackage{graphicx}
\usepackage{array}
\usepackage{multicol}
\usepackage{hhline}
\usepackage{xcolor}
\usepackage{perpage}
\MakePerPage{footnote}
\newcolumntype{C}[1]{>{\centering\arraybackslash}m{\dimexpr#1-2\tabcolsep\relax}}
\usepackage[colorlinks=true,
linkcolor=red,
urlcolor=blue,
citecolor=cyan]{hyperref}
\usepackage{ulem}
\usepackage[colorlinks=true,
linkcolor=red,
urlcolor=blue,
citecolor=cyan]{hyperref}
\title{Surface gravity in spherically symmetric collapsing stars}
\author{A. Sadeghi$^1$, F. Shojai$^{1}\footnote{Corresponding author: fshojai@ut.ac.ir}$\ , F. Bahmani$^{1}$\\$^1$Department of Physics, University of Tehran, P.O. Box 14395-547\\Tehran, Iran.}

\begin{document}
\maketitle
\begin {abstract}
Here we consider the generalized Oppenheimer-Snyder collapse of a star into a four-dimensional Einstein-Gauss-Bonnet black hole as well as a class of regular black holes labeled by the polytropic index of the stellar matter. We then analyze the nature of the horizon and the corresponding surface gravity outside and inside the star. The Hayward and Nielsen-Visser dynamical surface gravity are in agreement with the one resulting from the Killing vector of the outer static metric. However, these two definitions inside the star do not coincide with the Killing surface gravity outside the star when the star crosses the event horizon. This motivates us to study the surface gravity using Fodor's approach to have a unique surface gravity at the mentioned moment.
Then the extremality condition and the first law of thermodynamics are discussed at the trapping horizon of the star.
\end{abstract}
\section{Introduction}
Many efforts have been made to analyze the theory of general relativity (GR) by studying its solutions and their physical interpretations.
Black hole (BH) is one of the predictions of GR which is characterized by the existence of horizons, trapped surfaces, and singularities \cite{penrose}.

Although the naked singularities are covered by event horizons according to the cosmic censorship hypothesis \cite{pencos}, many efforts are made to introduce some BH geometries that are regular at all spacetime points. The first attempt was made by Einstein and Rosen by formulating the wormhole structure \cite{Visbook}. In the next attempt, Sakharov proposed $P = -\rho$ as the equation of state for the dense region of a star \cite{sakh}. One of the most important steps to remove the singularity was made by Bardeen by coupling the GR action to the nonlinear electrodynamics \cite{bar}. Then, Hayward introduced a regular BH solution with a fundamental length scale that prevents the formation of a singularity \cite{hay}.
Both the Bardeen and Hayward metrics which are of interest here, reduce to the deSitter metric at small distances while they  go to the Schwarzschild geometry asymptotically. A general class of regular BHs with deSitter core is introduced in \cite{our1} which reduces to the Hayward and Bardeen BHs for special values of the metric parameters. For a general review of various non-singular BHs see \cite{ansol}.

On the other hand, there are some extensions of GR which include functions of the curvature invariants in the action.
One of the important generalizations of GR is the Gauss-Bonnet (GB) theory, generalized by Lanczos \cite{lanczos} and Lovelock \cite{love}. This theory prevents the Ostrogradsky instability \cite{wood} by giving second-order field equations.
Although the consequences of GB gravity were supposed to be the same as those of GR in four-dimensional spacetime, recently it has been shown that a redefinition of the coupling constant of GB gravity can lead to new results even in four-dimensional spacetime \cite{Glav}. This theory is known as 4D Einstein-Gauss-Bonnet (4D-EGB) gravity. It should be noted that some authors have the opinion that the regularization method of \cite{Glav} is questionable and propose new regularization methods \cite{Fern},\cite{Henn}. Fortunately, even when other regularization methods are applied, the spherically symmetric BH solution of \cite{Glav} is still valid. For this solution, the metric tensor is regular in the center. However, the scalar curvature diverges at this point.

To get a physical intuition of a regular BH, the Oppenheimer-Snyder-Datt (OSD) collapse scenario \cite{opp} for a general class of regular BHs is studied in \cite{our1} and for the 4D-EGB BH geometry in \cite{our2}. In these works, the outer BH geometry is smoothly connected to the inner geometry of the star which is assumed to be spatially flat Friedmann-Robertson-Walker. It has been shown that the stellar matter could have uniform density and pressure to obtain a smooth transition at the star surface according to the Israel junction conditions \cite{our1}. In the generalized OSD collapse, the stellar matter will violate the strong energy condition (SEC) in the final stages of the collapse. This prevents the star from reaching the center in the case of a regular BH while it reduces the collapse velocity for 4D-EGB collapse and the star reaches the center with zero velocity.\\
As mentioned above, regular BHs are the solutions of non-vacuum Einstein equations but EGB BH is the solution of vacuum EGB gravity.
Therefore, EGB-gravity boundary conditions should be used for the latter. In \cite{refree}, the boundary conditions of charged anisotropic spherically symmetric stars in higher dimensional EGB gravity with a non-zero cosmological constant are obtained.
Using the modified field equations, the authors of \cite{refree} have generated a complete stellar model in EGB gravity and show that the radial pressure at the stellar surface should be zero, similar to the case of GR.
As expected, when the GB coupling constant tends to zero, the boundary conditions of N-dimensional GR are obtained.
Here we have described 4D-EGB gravity as GR with an effective stress-energy tensor containing higher-order curvature terms. Thus, we have used the Israel junction conditions of GR for the collapse to 4D EGB BH \cite{our2}. We will return to this point in Section \ref{sec2}.

In this paper, we want to investigate the nature of horizons in the generalized OSD collapse in terms of the ingoing and outgoing null geodesic congruences. The surface gravity (SG) is then calculated using the non-affinely parameterized geodesic on the trapping horizon (TH) \cite{nielsend}.

We also consider different approaches to SG, such as Hayward, Nielsen-Visser, and Fodor's approach.
We see that it is possible to obtain a unique SG on the horizon according to Fodor's approach. To do this, we use null vectors with normalization coefficients that are arbitrary functions of spacetime, the affinely parameterized ingoing null vector, and the condition of the uniqueness of the SG at the crossing moment.

We study the extremality condition for the outer and inner geometry and see that outside the star, SG vanishing and the coordinate-invariant definition of extremality are the same. But this is not the case for the TH inside the star.
We have also found the work term by writing the first law of thermodynamics using different
definitions of SG.

The outline of this paper is as follows:
In Section \ref{sec2}, we recall the OSD gravitational collapse for the 4D-EGB and the regular BHs. We consider different concepts of the TH and evaluate the SG in Section \ref{sec3}. In Sections \ref{sec4} and \ref{sec5}, we study the expansion parameter and the Lie derivative of the expansion parameter and then we evaluate the properties of the horizon in the outer and inner geometry, respectively. We study the SG of the event horizon (outside the star) and TH (inside the star) using different approaches in Section \ref{sec6}. In Section \ref{sec7}, following Fodor's approach, we obtain a condition that gives a unique SG at the horizon crossing time by choosing appropriate null vectors. Then, in Section \ref{sec8}, we study the extremality condition for the external and internal geometries. Section \ref{sec9} is devoted to obtaining the first law of thermodynamics for different approaches of Section \ref{sec3}. Section \ref{sec10} provides a summary and conclusion.

Throughout this paper, the signature of the metric tensor is assumed to be $(-, +, +, +)$. We use the natural units in which $G = c = 1$. All variables  are dimensionless with respect to the Schwarzschild radius. A dot denotes differentiation with respect to  the proper time of a freely falling particle on the surface of the star and a prime means the derivative with respect to the radial coordinate.
\section{OSD collapse into regular and 4D-EGB BHs}
\label{sec2}
Consider a particular class of static, asymptotically flat spherically symmetric BH geometries as
\begin{align}
\label{0}
ds^2 &= - \left(1 - \frac{1}{ r\omega( r)}\right) dt^2 + \left(1 - \frac{1}{ r\omega( r)}\right)^{-1} d r^2\nonumber\\
& + r^2 d \Omega^2
\end{align}
where the function $\omega( r)$ is defined by \cite{our1}
\begin{align}
\label{omegreg}
\omega_{\text{reg}}(r)&=\left(1+\beta r^{-3/n}\right)^n
\end{align}
for a regular BH with deSitter core and by \cite{our2}
\begin{align}
\label{omegegb}
\omega_{\text {EGB}}(r)&=\frac{32\pi\alpha }{r^3}\bigg[\sqrt{1+\frac{64\pi\alpha }{r^3}}-1\bigg]^{-1}
\end{align}
for a 4D-EGB BH. In the above expressions, $\beta$ is a free parameter, $n$ is the polytropic index of the stellar matter for the regular BH case, and $\alpha$ is the GB coupling constant. For $n=1$ and $n=3/2$, the metric \eqref{0} is reduced to the well-known Hayward \cite{hay} and Bardeen \cite{bar} metrics respectively while an arbitrary value of n gives a new family of static regular BHs with deSitter core. The location of the horizons of metric \eqref{0} can be obtained by setting $r \omega(r)=1$. In the regular case, there are two horizons for $\beta < 4/27$, the BH becomes extremal for $\beta = 4/27$, and there is no horizon for $\beta > 4/27$.
In the case of the 4D-EGB, there are two horizons for $\alpha<1/64\pi$. The value of $\alpha =1/64\pi$ corresponds to an extremal BH, while larger values of $\alpha$ lead to a horizonless geometry.

It is convenient to change the coordinates to the Painlev\'e-Gullstrand (PG) coordinates to construct a collapsing model for a star in the background geometry \eqref{0}. In terms of the PG coordinates we can write the general form of the spacetime metric as follows
\begin{align}\label{MMBB}
ds^2=-d\tau^{2}+(dr+f(r,\tau)d\tau)^2+r^2d\Omega^2
\end{align}
where $f(r,\tau)$ is given by
\begin{align}
\label{eq:LL}
\begin{cases}
-rH(\tau) \hspace{1cm} r\leq R\\
\sqrt{\frac{1}{r\omega(r)}}\hspace{1cm} r\geq R
\end{cases}
\end{align}
and $R$ is the radius of the star and the geometry inside the star is assumed to be described by the spatially flat FRW metric.
Given \eqref{eq:LL}, in order to have a smooth matching of the two geometries, we must have
\begin{align}
\label{feq}
\dot R=-\sqrt{\frac{1}{R\omega(R)}}
\end{align}
at the surface of the star. This means that the extrinsic curvature is the same on both sides of the star's surface and thus,  there is no stress-energy layer on the surface of the star according to the Israel junction conditions.

As mentioned before, here, we treat 4D EGB gravity as GR with an effective stress-energy tensor \cite{our2}.
In this way, the vacuum 4D EGB field equations can be written as 
\begin{align}\label{B}
G_{\mu\nu}+16\pi \alpha \mathcal{H}_{\mu\nu}=0
\end{align}
where $G_{\mu\nu}=R_{\mu\nu}-\frac{1}{2}g_{\mu\nu}R$ is the Einstein tensor and
\begin{align}\label{C}
\mathcal{H}_{\mu\nu}=&2(RR_{\mu\nu}-2R_{\mu\sigma}R^{\sigma}_{\nu}-2R_{\mu\sigma\nu\rho}R^{\sigma\rho}+R_{\mu\sigma\rho\delta}R^{\sigma\rho\delta}_{\nu})\notag\\
&-\frac{1}{2}(R_{\mu\nu\rho\sigma} R^{\mu\nu\rho\sigma}-4R_{\mu\nu}R^{\mu\nu}+R^2)g_{\mu\nu}
\end{align}
includes the higher-order terms of curvature. 
This picture is usually used for any other modified gravity theory, and so one can express the field equations in the Einstein form $G_{\mu\nu}=8\pi T_{\mu\nu}^{\text{eff}}$ where $T_{\mu\nu}^{\text{eff}}=T_{\mu\nu}^{\text{EGB}} +  T_{\mu\nu}^{\text{m}}$ is the effective stress-energy tensor. It includes the matter stress-energy tensor  $T_{\mu\nu}^{\text{m}}$  and the curvature terms that arise from the 4D EGB gravity, $T_{\mu\nu}^{\text{EGB}} =-2\alpha \mathcal{H}_{\mu\nu}$. 
This interpretation has been used extensively in f(R) gravity \cite{f(r)}, curvature-matter couplings \cite{curvmatt}, Weyl gravity \cite{manheim}, and in braneworlds \cite{sepangi}. In this way, the effective stress energy tensor includes higher order curvature terms responsible for violating the energy conditions. Moreover, this picture allows us to use the junction conditions of GR. Otherwise, one should consider the EGB boundary conditions \cite{refree}. Regarding the 4D EGB BH solution, if we insert \eqref{omegegb} into \eqref{0} and then substitute it into Einstein's equations, we get $\rho^{\text{EGB}}$ and the components of anisotropic pressure $p^{\text{EGB}}$. These must be added to the corresponding quantities of the star, $\rho^{\text{star}}$ and the components of $p^{\text{star}}$ in order to obtain the effective density and the different components of effective pressure of the star. This is done in \cite{our2} and it is shown that for a collapsing star, the radial component of the star's pressure would be zero at the surface of the star. This is similar to the result given by  \cite{refree}, but for a dynamically collapsing star.
 
\begin{figure}[h]
\centering
\includegraphics[width=2.35in]{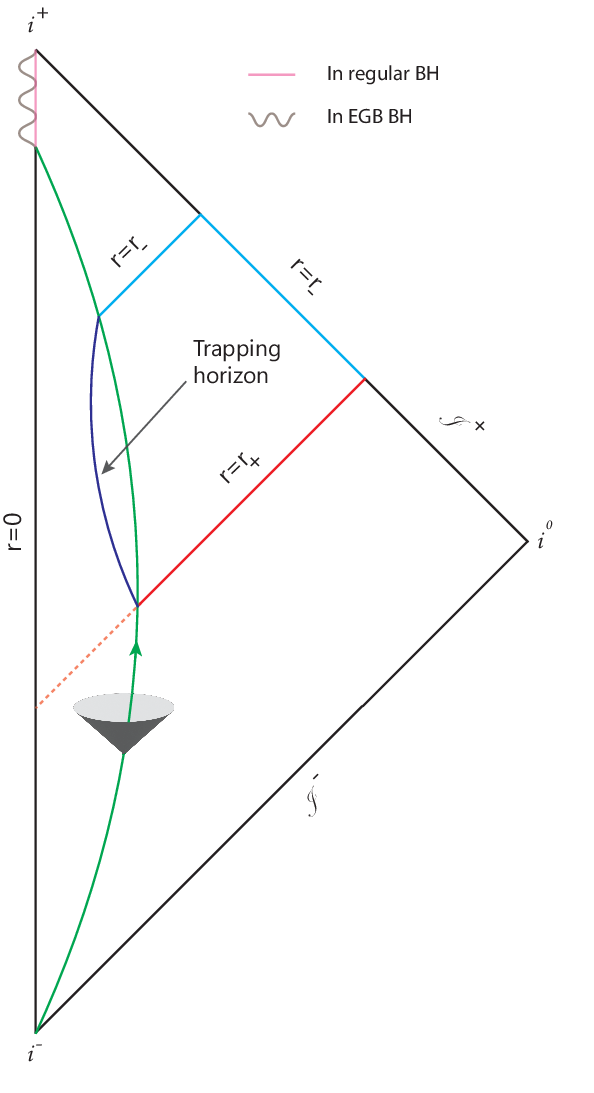}
\caption{Penrose diagram of a star collapsing into a BH with two horizons. The horizons are denoted by $r_+$ and $r_-$. The blue line is the time-like TH. It appears when the surface reaches $r_+$ and disappears when it reaches $r_-$. A straight line and a wavy line indicate the final state of the collapse for the singularity-free case in the regular metric
and with a singularity in the 4D-EGB case. }
\label{fig2}
\end{figure}

According to \eqref{feq}, each free particle on the surface of the star begins its free fall from rest at infinity and moves along a time-like radial geodesic.
Introducing the comoving radius as $R(\tau) = a(\tau)R_{\text{com}}$ and substituting it into \eqref{feq}, this equation has the form of Friedman's equation, $H^2=8\pi\rho/3$, if
\begin{equation}\label{den}
\rho(\tau)=\frac{3}{8\pi R^3\omega(R)}
\end{equation}
This gives the density of the star. It is given by
\begin{align}\label{den2f}
\rho_{\text{reg}}(\tau)=\frac{3}{8\pi}\frac{1}{(R(\tau)^{\frac{3}{n}}+\beta)^n}
\end{align}
for a regular BH and by
\begin{align}\label{den2}
\rho_{\text{EGB}}(\tau)=\frac{3}{256 \pi^2 \alpha}\left
(\sqrt{1+\frac{64 \pi \alpha }{ R(\tau)^3}}-1\right)
\end{align} for a 4D-EGB BH. Then, the pressure at the surface (and inside the star) can be read from the continuity equation, $\dot \rho+3H(\rho+P)=0$,
\begin{align}\label{p}
P(\tau)=\frac{\omega'(R)}{8\pi R^2\omega^2(R)}
\end{align}
which can be simplified to
\begin{align}\label{regp}
P_{\text{reg}}(\tau)=\frac{3}{8\pi}\frac{-\beta}{(R(\tau)^{\frac{3}{n}}+\beta)^{n+1}}
\end{align}
for a regular BH and
\begin{align}\label{SSlp}
\frac{8 \pi}{3}{P}_{\text{EGB}}(\tau)= \frac{1}{32\pi \alpha} - \big( \frac{1}{ R(\tau)^3} + \frac{1}{32 \pi \alpha} \big)\big( 1 + \frac{64 \pi \alpha}{R(\tau)^3}\big)^{-1/2}
\end{align}
for a 4D-EGB BH.
In the interior region of the star, the event horizon and TH can be found using the outgoing radial null geodesics. The time evolution of the horizons and the surface of the star is studied in detail in \cite{our1} and \cite{our2}. This is shown in the Penrose diagram of figure \ref{fig2}.
Assuming that the stellar matter is a perfect fluid, the equation of state can be derived from the geodesic equation of the stellar surface.
In the first case, this leads to polytropic matter, while in the second, a different form is obtained.
See equations (49) and (22) in \cite{our1} and \cite{our2} respectively.
\section{Trapping horizon and the surface gravity}
\label{sec3}
Consider a congruence of radial ingoing and outgoing
null geodesics with tangent fields $n^a$ and $\ell^a$ which are orthogonal to the 2-spheres of symmetry and satisfying the cross normalization $n^a \ell_a =-1$.
The corresponding expansion parameters are denoted by $\theta_\ell$ and $\theta_n$. The expansion parameter of $\ell^a$ is defined as
\begin{align}
\label{thet}
\theta_\ell = g^{ab} \nabla_a \ell_b + n^a \ell^b \nabla_a \ell_b + \ell^\alpha n^b \nabla_a \ell_b
\end{align}
and $n\leftrightarrow \ell$ is substituted to obtain $\theta_n$, the expansion parameter of the ingoing null vector field. A TH \cite{Hayward1994} is a hypersurface that can be foliated by 2-surfaces such that $\theta_\ell\big|_\text{h}=0$, $\theta_n\big|_\text{h}\neq0$ and $n^a \nabla_a \theta_\ell\big|_\text{h} \neq 0$. A TH is called outer if $n^a \nabla_a \theta_\ell\big|_\text{h} < 0$, inner if $n^a \nabla_a \theta_\ell\big|_\text{h} > 0$ , future if $\theta_n\big|_\text{h}<0$ , and past if $\theta_n\big|_\text{h}>0$. A marginally trapped tube T (MTT) \cite{10JFT} is a hypersurface that is foliated by 2-surfaces S, called marginally trapped surfaces (MTS), such that $\theta_\ell\big|_T=0$ and $\theta_n\big|_T<0$. If a MTT is spacelike (timelike) everywhere, it is called a dynamical horizon (time-like membrane)\footnote{A short and useful definition of the different quasi-local horizons is given in \cite{Booth}, Table 1.}.

For THs, the change in the area
of the horizon can be related to the local value of $T_{ab} \ell^a\ell^b$ \cite{Hayward1994}. From the Raychaudhuri equation with $\theta_\ell=0$ and assuming that $l^\alpha $ is hypersurface orthogonal, one has
\begin{align}
\label{area}
r^{b} \nabla_b\delta A = \frac{\zeta \theta_n \delta A}{n^{b} \nabla_b \theta_\ell} \left( \sigma_{ab}\sigma^{ab} + R_{ab} \ell^a\ell^b\right)
\end{align}
where $r^a $ is a vector that is tangent to the TH and normal to the foliation by 2-surfaces, $\delta A $ is an infinitesimal circle drawn on the spacelike 2-surface,  $\sigma_{ab}$ is the shear tensor corresponding to the
outgoing null vector  and $\zeta$ is a scalar field at the TH. Assuming that $\zeta >0$\footnote{This can be satisfied by fixing the orientation of $r^a $.}, since
$\theta_n<0$ and $n^{b} \nabla_b \theta_\ell<0$ for a future outer TH (FOTH) and $n^{b} \nabla_b \theta_\ell>0$ for a future inner TH (FITH), the sign of the area change is determined by the null energy condition\footnote{Note that the location of the TH and the sign of $n^{b} \nabla_b \theta_\ell$ are independent of the coefficient functions of the  ingoing and outgoing null vectors.}. The sign is negative for FITH and positive for FOTH if the null energy condition is satisfied.
The existence of a decreasing area horizon is reported for a
time-like membrane in some shell collapse models \cite{Booth2006}.

In stationary spacetime, the SG is a measure of the non-affinity of the Killing vector at the Killing horizon, where it becomes null.
The SG of a non-Killing horizon in a spherically symmetric BH is usually defined as the inaffinity of the outgoing radial null geodesic $\ell ^a$ at the horizon. Several proposals have been made to fix the normalization of $\ell ^a$. In this paper, we focus on three approaches\footnote{There are several other definitions that we will not mention here. They do not give the correct value of SG in stationary spacetime. Or they use a special normalization of the outgoing null vector after selecting the coordinate system. We refer the interested reader to \cite{Hayward1994}, \cite{Collins:1992eca} and \cite{Abreu:2010ru}.}. 

The first one is the proposal of Fodor et al \cite{Fodor:1996rf} which is based on an affinely parametrized ingoing null geodesic $n^a$ whose asymptotic form is such that $\xi^a n_a=-1$ in an asymptotically flat geometry where $\xi^a$ is the asymptotic time-translational Killing vector. The cross-normalization condition $n^a \ell_{a }=-1$ leads to
\begin{align}\label{KF}
\kappa_{\mathrm{F}}=-n^{a}\ell^{b} \nabla_{b}\ell_{a}
\end{align}
Following Fodor's method, there will be no unknown degrees of freedom left in the ingoing null vector field and hence in the outgoing corresponding one, when the SG is read out in advanced Eddington-Finkelstein coordinates.  So the SG is completely determined \cite{Fodor:1996rf}. However, this is not true in PG coordinates as we will see in Section \ref{sec7}.

The second is Hayward's proposal \cite{Hayward:1997jp}. It is based on the definition of the SG of spherically symmetric spacetime in terms of the Kodama vector $K^{a}$ \cite{Kodama:1979vn}.
Here we want to explain the Kodama SG in detail. This is because there is a small error in this definition that has been repeated in the literature and needs to be clarified. Assume that the spacetime metric is given by
\begin{equation}\label{h}
ds^{2}=h_{ab}dx^a dx^b+{\cal R}^2 d\Omega^{2}_{(2)} , \quad a,b=0,1
\end{equation}
 where ${\cal R}$ is the areal radius and $\epsilon_{ab}$ is the volume form of $h_{ab}$.
The Kodama vector is defined as
\begin{equation}
\label{koda1}
K^a \equiv \epsilon^{ab} \nabla_b{\cal R}
\end{equation}
To find the SG, it is necessary to calculate
\begin{align}\label{HoH}
\frac{1}{2}g^{a b}&K^{c}(\nabla_{c}K_{a}- \nabla_{a}K_{c})\nonumber\\
&=\frac{1}{2}g^{a b}\nabla_{d}{\cal R}(\epsilon ^{c d}\epsilon _{a e}\nabla_{c}\nabla^{e}{\cal R}-\epsilon ^{c d}\epsilon _{c e}\nabla_{a }\nabla^{e}{\cal R})\nonumber \\&=
\frac{1}{2}g^{a b}\nabla_d {\cal R}(\delta^{d}_{a} \square {\cal R})\nonumber= \frac{1}{2}(\nabla^{b}{\cal R})\square {\cal R}\nonumber\\
&\equiv \kappa_{_\mathrm{H}}\epsilon ^{b c}\epsilon _{c d}\nabla^{d}{\cal R}
= \kappa_{_\mathrm{H}}\epsilon ^{b c}K_{c}
\end{align}
where in the last line the Hayward definition of SG
\begin{align}\label{SGH}
\kappa_{_\mathrm{H}}=\frac{1}{2}\square _h{\cal R}
\end{align}
and \eqref{koda1} are used.
We see that $\epsilon ^{b c}$ appears on the right side of \eqref{HoH}.  The Kodama equation, which appears in \cite{nielsend} and \cite{faraoni} as well as in other literature, omits $\epsilon ^{bc}$ and uses $K_c$ on the right hand side. 
However, the presence of $\epsilon ^{b c}$ on the right side of \eqref{HoH} is expected since it is a consequence of the antisymmetrization of the left side of \eqref{HoH}. It should be noted that the Kodama vector and the value of SG remain unchanged by this correction.
Both Hayward and Fodor SGs are valid at all points in spacetime however we are usually interested in their values at a FOTH.

The third definition is that of Nielsen and Visser \cite{vissern} who use the PG coordinates. 
Consider the general metric form in PG coordinates
\begin{align}\label{viss}
ds^2 &= - [c(r,\tau)^2-v(r,\tau)^2]d \tau^2 + 2 v(r,\tau) d r d \tau
+ d r^2\nonumber \\&+ r^2  d\Omega^2.    
\end{align} 
The authors of \cite{vissern} define the outgoing and ingoing radial null vectors as follows
\begin{equation}\label{gggg}
\ell^a={\left(\vphantom{\Big{|}}1,c(r,\tau)-v(r,\tau),0,0\right)\over c(r,\tau)},
\end{equation}
\begin{equation}\label{gggg1}
n^a={\left(\vphantom{\Big{|}}1,-c(r,\tau)-v(r,\tau),0,0\right)\over c(r,\tau)},
\end{equation}
In the above, the normalization factors are chosen so that the null vectors satisfy two conditions. The first is the cross-normalization  relation, $n^{a}\ell_{a}=-2$ and the other is related to the SG of the TH. This function  must be consistent with what is found in the  first law of BH thermodynamics written by partial differentiation of the Misner-Sharp mass \cite{Misner} 
\begin{equation}\label{def}
m(\tau,r)=r v^2(\tau,r)/{2 c^2(\tau,r)}
\end{equation}
with respect to  proper time. Doing this and  then evaluating the result at the evolving horizon $2m(r,\tau)=r$, we get
\begin{equation}
 \label{E:mdot}
 \dot m(r_\text{h}(\tau),\tau) = {[1- 2 m'(r_\text{h}(\tau),\tau)]\over 2}\;  \dot r_\text{h}(\tau).
 \end{equation} 
This can be rewritten in terms of the area of the evolving horizon  $A_h = 4\pi r_\text{h}^2$ as
\begin{equation}\label{semi f}
 \dot m(r_\text{h}(\tau),\tau)
 = {1\over 8\pi} {[1- 2 m'(r_\text{h}(\tau),\tau)] \over 2 r_\text{h}(\tau)} \dot
{A}_\text{h}(\tau).
 \end{equation}
This is the first law of BH thermodynamics if 
\begin{equation}\label{SGNV}
\kappa_{\text{NV}}(\tau) = {1-2m'(r_\text{h}(\tau),\tau)\over2r_\text{h}(\tau)},
\end{equation}
The authors of \cite{vissern} define the SG based on the inaffinity of the outgoing null vector field and as we have mentined, fix its normalization to obtain the SG \eqref{SGNV}.
This gives \eqref{gggg} and \eqref{gggg1}.
We will return to this point in Section \ref{sec9}. 

The key concept to note is that the Hayward and Nielsen-Visser definitions of SG are not the same as Killing SG in static spacetime when $g_{tt}g_{rr}\neq{-1}$ in Schwarzschild-like coordinates
\cite{nielsend}, \cite{Pielahn:2011ra}. Thus, these definitions are identical to the Killing SG for the static outer geometry of the star, given by \eqref{0}.
If the FRW geometry inside the star is reduced to the deSitter geometry where the Hubble parameter is independent of time, the above condition on the metric components holds. 
Otherwise, the Hayward and Nielsen-Visser SGs lead to different results, as we will see in Section \ref{sec6}.

The next two sections discuss the properties of the outer and inner geometry of the collapsing star. Then we will use the three definitions \eqref{KF}, \eqref{SGH}, and \eqref{SGNV} above to calculate the SG in Section \ref{sec6}. 
\section{External geometry of the star}
\label{sec4}
Outside the star, the geometry is static. Therefore, the event horizon coincides with the OTH. The radius of the TH of metric \eqref{0} is given by $r_{\pm} \omega(r_{\pm}) =1$ which gives
\begin{align}
\label{rhreg}
r_{\pm \text{(reg)}}&= \left(\frac{1}{3} + \frac{2}{3}\cos\big(\frac{\pi}{3} \mp \frac{1}{3} \cos^{-1} [\frac{27 \beta}{2} -1]\big)\right)^n\\
\label{rhegb}
r_{{\pm}{\text{(EGB)}}} &=\frac{1}{2}\left(1 \pm \sqrt{1- 64\pi \alpha}\right)
\end{align}
Here we have considered the case with two horizons (i.e. $\beta<4/27$ and $\alpha<1/64\pi$).  
Using metric \eqref{0} and cross-normalization $n^a\ell_a=-1$, a specific set of radial null vectors is as follows
\begin{align}\label{ln}
\ell^b_{\text{out}}& = (1, 1- \sqrt{\frac{1}{ r \omega(r)}} ,0 ,0)\\
n^b_{\text{out}}& =\frac{1}{2} (1, -1- \sqrt{\frac{1}{r \omega( r)}} ,0,0)\label{nl}
\end{align}
By a simple calculation, the expansion parameter of $n$ and $\ell$ using equation \eqref{thet} can be obtained as
\begin{align}\label{l}
\theta_{\ell,\text{out}} &= \frac{2}{r}\left( 1- \sqrt{\frac{1}{r \omega(r)}}\right)\\
\theta_{n,\text{out}} &= -\frac{1}{r}\left( 1+ \sqrt{\frac{1}{r \omega(r)}}\right)
\end{align}
where at the horizons 
\begin{align}
\theta_{\ell,\text{out}}\big|_{r_{\pm}} =0\,\,\,\,\,\,\,\,\, \theta_{n,\text{out}}\big|_{r_{\pm}} =-\frac{2}{r}<0
\end{align}
which means  that $r_{\pm}$ are MTTs. 
Furthermore, using \eqref{0} it can be shown explicitly that 
\begin{align}
\label{ndelh}
n^{b}_{\text{out}} \nabla_b \theta_{\ell,\text{out}} \big|_{{r_\pm}} =- \frac{1}{ r_{\pm}^2} \left( 1 + r_{\pm}^2\omega^\prime ( r_{\pm})\right)
\end{align}
The sign of the expression \eqref{ndelh} depends on the sign of $1+ r_{\pm}^2 \omega^\prime ( r_\pm)$. For regular BH
\begin{align}
1+r_{\pm \text{(reg)}}^{2} \omega^\prime_{\text{(reg)}} ( r_{\pm})&=1-3 \beta  r_{{\pm}{\text{(reg)}}}^{-\frac{n+2}{n}}\nonumber\\
\label{rom1}
= &1- 3^{4-n}\beta   X_{\pm}^{n-3} \left(27 \beta X_{\pm}^{-3}+1\right)^{n-1}
\end{align}
where $X_+=1+2 \cos \left(\frac{1}{3} \cos ^{-1}[1-\frac{27 \beta }{2}]\right)$ and $X_-=1-2 \sin \left(\frac{1}{3} \sin ^{-1}[1-\frac{27 \beta }{2}]\right)$. The expression \eqref{rom1} is positive in the outer horizon $r_ +$ (so \eqref{ndelh} is negative) and negative in the inner horizon $r_ -$ for all allowed values of  $\beta$ and $n$.
The same result is obtained for the 4D-EGB BH for all allowed values of $\alpha$.
This can be seen from
\begin{align}
\label{rom2}
1+ {r_{{{\pm}}{\text{(EGB)}}}^{2}} {{\omega^{\prime}}_\text{(EGB)}}( r_{\pm})=1- \frac{48 \pi \alpha }{r_{{\pm}\text{(EGB)}}^{2}}\big(\frac{64 \pi \alpha }{r_{{\pm}\text{(EGB)}}^{3}}+1\big)^{-\frac{1}{2}}
\end{align}
and substituting $r_{\pm\text{(EGB)}}$ from \eqref{rhegb} in the above. Therefore, as expected,  $r_+$ is the FOTH for both BHs.
\section{Internal geometry of the star}
\label{sec5}
Inside the star, spacetime is dynamical and so, the definition of the event horizon turns out to be substantially different from the  quasi-local horizon definitions mentioned in Section \ref{sec3}. Using \eqref{MMBB}, and cross-normalization $n^a\ell_a=-1$, a specific choice of the radial null vectors  is given by 
\begin{align}\label{llnn}
\ell^a_{\text{in}} &= (1, 1+ r H ,0 ,0)\\
n^a_{\text{in}} &= \frac{1}{2} (1, -1 + r H,0,0)\label{nnll}
\end{align}
It is then easy to show that in this region
\begin{align}
\theta_{\ell,\text{in}} &= \frac{2}{r}\left( 1+ r H\right)\\
\theta_{n,\text{in}} &= -\frac{1}{r}\left( 1-r H\right)
\end{align}
For $r=-1/H$,
\begin{align}
\theta_{\ell,\text{in}}\big|_\text{h} =0, \,\,\,\,\,\,\,\,\, \theta_{n,\text{in}}\big|_\text{h}<0
\end{align}
So, it is a TH. On this surface, the Lie derivative of the expansion parameter becomes
\begin{align}\label{gg}
n^a \nabla_a \theta_\ell\big|_{\text{h},\text{in}} =\dot{H}+2H^2
\end{align}
Using Friedmann's equations, $H^2=8\pi\rho/3$ and $\dot{H}=-4\pi (\rho+P)$, we see that
\begin{align}
\dot{H}+2H^2=\frac{4\pi}{3}(\rho-3P)
\end{align}
The pressure inside and on the surface of the star is negative, according to (\ref{regp}) and (\ref{SSlp}), for regular and 4D-EGB BHs. Therefore, expression (\ref{gg}) is positive and the TH is a FITH. 
We have already shown that the total energy-momentum tensor inside the star satisfies the NEC \cite{our1}, \cite{our2}.
Therefore, the area of the horizon inside the star can be reduced even if the null energy condition is satisfied, according to \eqref{area}.
This is consistent with the result of the horizon evolution equation obtained in \cite{our1} and other collapse models \cite{nielsen1}.
\section{Surface gravity}
\label{sec6}
In this Section, we will calculate the SG for the inside and outside of the star. As mentioned in Section \ref{sec3}, the definition of Hayward's SG in terms of Kodama's vector and the definition of Nielsen-Visser's SG are not necessarily the same. To clarify this point, let us consider the following general metric in PG coordinate \cite{Pielahn:2011ra}
\begin{align*}
ds^2=-\sigma(t,r)^2dt^2+(dr+\sqrt{\frac{2m}{r}}dt)^2+r^2 d\Omega^2
\end{align*}
where $\sigma(t, r)$ is equal to one if $g_{tt} g_{rr} = -1$.
In the static case, the Killing SG is:
\begin{align}\label{sh2}
\kappa_{\text{killing}}=\frac{\sigma(t,r)}{4m}(1-2m')
\end{align}
For the dynamical case, according to the definition of SG by Hayward and Nielsen-Visser, we obtain
\begin{align}\label{sh3}
\kappa_{\text{H}}=\frac{1}{4m}(1-2m')+\frac{\dot{m}}{4\sigma(t,r)}
\end{align}
\begin{align}\label{sh3}
\kappa_{\text{NV}}=\frac{1}{4m}(1-2m')
\end{align}
Thus, the two definitions give the Killing SG for static spacetime, such as the outer region of the star. However, they are not the same in the dynamical case due to the existence of $\dot{m}$, such as the inner region of the star, see \eqref{eq:LL}.
Below, we will see that these two definitions are identical for the TH inside the star if $\dot{H}=0$, the de Sitter geometry. 

Here, both the inner and outer regions of the star are represented by the PG coordinates.
Therefore, we use the general form of the spacetime metric as \eqref{MMBB}.
First, we use Hayward's method to calculate the SG. Substituting the two-dimensional part of metric \eqref{MMBB} into Hayward formula \eqref{SGH}, we get
\begin{align}
\kappa_{\mathrm{H}}=\frac{1}{2}(\dot{f}-2ff')
\end{align}
Therefore,
\begin{align}\label{gru}
\kappa_{\mathrm{H}}^{\text{(out)}}\big|_{r=r_+}&=\frac{r\omega'(r)+\omega(r)}{2(r\omega(r))^2}\big|_{r=r_+}\nonumber\\
&=\frac{r_{+}\omega'(r_{+})+\omega(r_{+})}{2}\\
\label{gri}
\kappa_{\mathrm{H}}^{\text{(in)}}\big|_{r=r_{\text{TH}}(\tau)}&=\frac{-r}{2}(\dot{H}(\tau)+2H(\tau)^2)\big|_{r=r_{\text{TH}}(\tau)}\nonumber\\
&=\frac{\dot{H}(\tau)+2H(\tau)^2}{2H(\tau)}
\end{align}
where it can be evaluated at each constant time slice of the TH\footnote{Although the horizon $-1/H$ is a FITH, we will use TH here for the sake of brevity.}.
According to Fig \ref{fig2}, the TH is formed when the surface of the star crosses $r_+$. At this moment, $\tau_c$, i.e. the crossing time, the radius of the star is $R_c$ where
\begin{align}\label{cr}
R_\text{c}=r_+=-1/H(\tau_\text{c})=1/\omega(R_\text{c})
\end{align}
from \eqref{eq:LL} and a natural question is, are the SGs of \eqref{gru} and \eqref{gri} equal at this moment?
To answer this question, we must express the inner SG \eqref{gri} in terms of the density and pressure of the star by using Friedmann's equations and then use equations \eqref{den} and \eqref{p} to write \eqref{gri} in terms of the $\omega$ function and its derivative at the star's surface. Performing this calculation, we find that
\begin{align}\label{o1}
\kappa_{\mathrm{H}}^{\text{(in)}}\big|_{r=r_{\text{TH}}(\tau_\text{c})}=\frac{R_\text{c}\omega'(R_\text{c})-\omega(R_\text{c})}{4}
\end{align}
at crossing time. This is different from (\ref{gru}) where its value at this moment is
\begin{align}\label{hayhay}
\kappa_{\mathrm{H}}^{\text{(out)}}\big|_{r=R_\text{c}}=\frac{R_\text{c}\omega'(R_\text{c})+\omega(R_\text{c})}{2}
\end{align}
Now, it is necessary to compare the two metrics \eqref{MMBB} and \eqref{viss} in order to find the SG at crossing time using the Nielsen-Visser method.
Remembering the definition of the Misner-Sharp mass \eqref{def}, we then obtain from \eqref{SGNV} on the horizon where $f=1$,
\begin{align}
\kappa_{\text{NV}}=-f'
\end{align}
which gives
\begin{align}\label{asd}
\kappa_{\text{NV}}^{\text{(in)}}\big|_{r=r_{\text{TH}}(\tau)}&=H(\tau)\\
\label{poi}
\kappa_{\text{NV}}^{\text{(out)}}\big|_{r=r_+}&=\frac{r\omega'(r)+\omega(r)}{2(r\omega(r))^{3/2}}\big|_{r=r_+}\nonumber\\
&=\frac{r_{+}\omega'(r_{+})+\omega(r_{+})}{2}
\end{align}
These can be simplified at the crossing time  to
\begin{align}\label{o2}
\kappa_{\text{NV}}^{\text{(in)}}\big|_{r=r_{\text{TH}}(\tau_\text{c})}&=-\omega(R_\text{c})\\
\label{visvis}
\kappa_{\text{NV}}^{\text{(out)}}\big|_{r=R_\text{c}}&=\frac{R_\text{c}\omega'(R_\text{c})+\omega(R_\text{c})}{2}
\end{align}
 The result of \eqref{poi} shows that the Nielsen-Visser SG of the outer horizon is the same as \eqref{gru}, as expected. So it is obvious that at the crossing time, when the surface of the star reaches the outer event horizon, it is reduced to \eqref{hayhay}. This is because the geometry outside the star is static and has a time-like Killing vector. According to the standard definition of the Killing SG for Killing horizons, a simple calculation gives \eqref{gru} as the Killing SG.
 
In general, however, the SG evaluated on the $r_+$ and TH of the star will not necessarily coincide at the crossing time. This is because the metric inside the star is inherently dynamic for which there is no unique definition of SG. This point can be easily seen from \eqref{o1} and \eqref{o2}.
\section{A unique surface gravity at crossing time}
\label{sec7}
Now we want to study SG using Fodor's approach. In this approach, the ingoing radial null rays satisfy the affinely parametrized geodesic equation\footnote{Note that the ingoing radial vectors (\ref{nl}) and (\ref{nnll}) are not affinely parameterized.}. To obtain the same inner and outer surface gravities at the crossing time, we use some arbitrary normalizing functions in the null vectors and then specify these functions. 
The outgoing and ingoing radial null vectors for metric (\ref{MMBB}) are
\begin{align}
\label{pu}
\ell^{a}&=C(r,\tau)(1,1-f,0,0)\nonumber\\
n^{a}&=D(r,\tau)(1,-1-f,0,0).
\end{align}
Assume that the geodesic equation for  $n^a$ is affinely parameterized. Then we obtain the following equation for $D(r,\tau)$:
\begin{align}
&D\dot{D}+D(-1-f(r,\tau))D'+\Gamma^{0}_{00}D^2\nonumber\\
\label{equ}
 &+2\Gamma^{0}_{10}D^2(-1-f(r,\tau))+\Gamma^{0}_{11}D^{2}(-1-f(r,\tau))^{2}=0
\end{align}
where $\Gamma^{0}_{00}=-f^2f'$, $\Gamma^{0}_{10}=-ff'$, $\Gamma^{0}_{11}=-f'$ are the Christoffel symbols of metric \eqref{MMBB}. Substituting these symbols into \eqref{equ} yields
\begin{align}\label{GH}
\dot{D}-\left(1+f(r,\tau)\right)D'-f'(r,\tau)D=0.
\end{align}
with solution 
\begin{align}
D_{+}(r>R)&=\frac{E}{1+ \left(r \omega(r)\right)^{-1/2}} \nonumber\\
&\times\exp{\left[\gamma \big(\tau+\int \frac{dr}{1+ \left(r \omega(r)\right)^{-1/2}} \big)\right]}\\
D_{-}(r<R)&=E'\exp{\bigg[\gamma'\big (r e^{-\int H(\tau) d\tau} } \nonumber\\
&{+\int e^{-\int^{\tau} H(\tau') d\tau'} d\tau\big ) -\int H(\tau) d\tau\bigg]}
\end{align}
where $E$, $E'$, $\gamma$ and $\gamma'$ are integration constants. As mentioned in Section \ref{sec3} for an asymptotically flat geometry with the time-translational Killing vector $\xi$, 
we require that $\xi^{a}n_{a}=-1$ at spatial infinity, therefore
\begin{align}
\label{cc}
\lim_{r\to\infty} D(r,\tau)=1. 
\end{align}
This means that
\begin{align}
D_{+}(r>R)&=\frac{E}{1+ \left(r \omega(r)\right)^{-1/2}}\label{FFR}\\
 \label{KKJ}
D_{-}(r<R)&=E' \exp{\left(-\int H(\tau) d\tau\right)}
\end{align} 
From the  zeroth component of the outgoing radial null geodesic, $\ell^{a}\nabla_{a}\ell^{b}=\kappa_{F}\ell^{b}$, we have
 \begin{align}\label{pp}
 \kappa_{\mathrm{F}}=\dot{C}+C'-C f'(r,\tau)-C' f(r,\tau).
 \end{align} 
Now, if we use the cross-normalization $\ell^{a}n_{a}=-1$, we have $2CD=1$ and equation (\ref{pp}) becomes 
 \begin{align}\label{uu}
 \kappa_{\mathrm{F}}=\frac{-1}{2{D^2}}(\dot{D}+D'+D f'(r,\tau)-D' f(r,\tau))
 \end{align}   
This gives the SG at $r_+$ and TH using (\ref{MMBB}), (\ref{FFR}) and (\ref{KKJ})
\begin{align}\label{ufu}
 \kappa_{\mathrm{F}}^{\text{(in)}}\bigg|_{r=r_{\text{TH}}(\tau)}&=\frac{H}{E'}\exp{\left({\int_{\tau_c}^{\tau} H(\tau') d\tau'}\right)\bigg|_{r=r_{\text{TH}}(\tau)}}\nonumber\\
&=\frac{H(\tau)}{E'}\frac{R(\tau)}{R_\text{c}}\bigg|_{r=r_{\text{TH}}(\tau)}=\frac{-1}{E'R_\text{c}}\frac{R(\tau)}{r_{{\text{TH}}}(\tau)}\\
\label{ufw}
 \kappa_{\mathrm{F}}^{\text{(out)}}\bigg|_{r=r_+}&=\frac{r\omega'(r)+\omega(r)}{2E(r\omega(r))^2}\bigg|_{r=r_+}
 \end{align}
where in the second line of \eqref{ufu}, the definition of the Hubble parameter $H=\dot{R}(\tau)/R(\tau)$ is used. Also, we choose $E = 1$ to match \eqref{ufw} with the SG given by  \eqref{gru} and \eqref{poi} for the outer metric. 
\begin{table*}[h!]
\caption{SG of horizons of a collapsing star using  different approaches. Outside the star, all of these approaches lead to the same result at the horizon. The TH appears at $\tau_c$ and then vanishes at $\tau_e$. The variables are dimensionless with respect to the Schwarzschild radius.}
\begin{center}
\renewcommand{\arraystretch}{1.1} 
\setlength{\tabcolsep}{0.2cm}
\label{table}
\begin{tabular}{ |C{3cm}||C{3.9cm}|C{3.4cm}| C{4cm} || C{3.5cm} | }
\hline
Approaches& Hayward & Nielsen-Visser & Fodor & Killing (outside the star)  \\[2mm]\hline
&&&&\\[0.2mm]
SG: evaluation at the event horizon outside the star    &    $\frac{r\omega'(r)+\omega(r)}{2\left(r\omega(r)\right)^{2}}\big|_{r_{+}\omega(r_{+})=1} $&  $\frac{r\omega'(r)+\omega(r)}{2\left(r\omega(r)\right)^{3/2}}\big|_{r_+\omega(r_+)=1}$ & $\frac{r\omega'(r)+\omega(r)}{2  (r\omega(r))^2}\big|_{r_+\omega(r_+)=1}$&$\frac{r \omega^\prime(r) + 1/r}{2}\big|_{r_+\omega(r_+)=1}$\\[7mm]\hhline{|----|~}
&&&&\\[0.2mm]
SG: evaluation at the TH  inside the star &$\frac{1}{2 H}(\dot H + 2 H^2)\big|_{\tau_\text{c}<\tau<\tau_\text{e}}$ &$ H\big|_{\tau_\text{c}<\tau<\tau_\text{e}}$ &$-\frac{1+{r_+^2}\omega'(r_+)}{2} \frac{\dot{R}(\tau)}{R_\text{c}}\big|_{\tau_\text{c}<\tau<\tau_\text{e}} $&-\\[7mm]
SG: inside the star at crossing time  &$\frac{1}{4} \left( R_\text{c}\omega^\prime(R_\text{c}) - \frac{1}{R_\text{c}}\right)$ & $ -\omega(R_\text{c})$ &$\frac{1+{r_+^2}\omega'(r_+)}{2}\omega(R_\text{c})$&$\frac{1}{2} \left( R_\text{c} \omega^\prime(R_\text{c}) + \frac{1}{R_\text{c}}\right)$ \\[7mm]
\hline
\end{tabular}
\end{center}
\end{table*}
\begin{table*}[h!]
\caption{Different SGs for regular and 4D-EGB BHs. The variables are dimensionless with respect to the Schwarzschild radius.}
\begin{center}
\renewcommand{\arraystretch}{1.1} 
\setlength{\tabcolsep}{0.2cm}
\label{table2}
\begin{tabular}{ |C{2.35cm}||C{3.9cm}|C{4.9cm}| C{3.9cm} || C{3.6cm} | }
\hline
Different SGs & Hayward & Nielsen-Visser & Fodor & At the event horizon\\[0.3mm]\hline
&&&&\\[0.1mm]
Hayward BH ($n=1$ in \eqref{omegreg})&$\frac{1}{2}\frac{r \left(r^3-2 \beta \right)}{ \left(r^3+\beta \right)^2}\big|_{r_{+}}$  & $\frac{1}{2} \frac{r^3 - 2 \beta}{\left(r^3+\beta \right)^{3/2}}\big|_{r_{+}} $   &$\frac{1}{2}\frac{r \left(r^3-2 \beta \right)}{ \left(r^3+\beta \right)^2}\big|_{r_{+}}$& $\frac{1}{2 r_{+}}+\frac{3}{2r_{+}} \left(r_{+}-1\right)$\\[7mm]
Bardeen BH ($n=3/2$ in \eqref{omegreg})&$\frac{1}{2} \frac{r(r^2 - 2 \beta)}{(r^2 + \beta)^{5/2}}\big|_{r_{+}}$& $\frac{1}{2} \frac{r^2 - 2\beta}{(r^2 + \beta)^{7/4}}\big|_{r_{+}}$ &$\frac{1}{2} \frac{r(r^2 - 2 \beta)}{(r^2 + \beta)^{5/2}}
\big|_{r_{+}}$&  $\frac{1}{2 r_{+}}+\frac{3}{2r_{+}} \left(r_{+}^{2/3}-1\right)$\\[7mm]
4D EGB BH  &$\frac{r^3 \left(\sqrt{\frac{64 \pi  \alpha }{r^3}+1}-1\right)-16 \pi  \alpha }{32 \pi  \alpha  r^2 \sqrt{\frac{64 \pi  \alpha }{r^3}+1}}\big|_{r_{+}}$ & $\frac{\left(3-\sqrt{\frac{64 \pi  \alpha }{r^3}+1}\right) \left(\sqrt{\frac{64 \pi  \alpha }{r^3}+1}-1\right)^{\frac{1}{2}}}{16 \sqrt{2 \pi \alpha}  \sqrt{\frac{64 \pi  \alpha }{r^3}+1}}\big|_{r_{+}}$&$\frac{r^3 \left(\sqrt{\frac{64 \pi  \alpha }{r^3}+1}-1\right)-16 \pi  \alpha }{32 \pi  \alpha  r^2 \sqrt{\frac{64 \pi  \alpha }{r^3}+1}}\big|_{r_{+}}$&  $  \frac{1}{2r_{+}} + \frac{3}{2 r_{+}} \frac{(r_{+}-1)}{|r_{+}-2|} $ \\[5mm]
\hline
\end{tabular}
\end{center}
\end{table*}
Evaluating \eqref{ufu} and \eqref{ufw} at the crossing time gives:
\begin{align}\label{uu}
\kappa_{\mathrm{F}}^{\text{(in)}}\big|_{r=r_{\text{TH}}(\tau_\text{c})}&=\frac{H(\tau_\text{c})}{E'}\\
\label{uw}
\kappa_{\mathrm{F}}^{\text{(out)}}\big|_{r=R_\text{c}}&=\frac{R_\text{c}\omega'(R_\text{c})+\omega(R_\text{c})}{2}
\end{align}
Now, we find the coefficient $E'$ with the condition that \eqref{uu} and \eqref{uw} are equal at the crossing time. At this moment, using  \eqref{den} and \eqref{p}, the right-hand side of \eqref{uw} can be rewritten in terms of the density and pressure of the star and thus from Friedmann's equations in terms of the Hubble parameter and its derivative, i.e.
\begin{align}
\frac{R_\text{c}\omega'(R_\text{c})+\omega(R_\text{c})}{2}=\frac{4\pi}{3}R_\text{c}\left(\rho(\tau_\text{c})+3P(\tau_\text{c})\right)\nonumber\\
=-R_\text{c}(\dot{H}+H^2)\big|_{\tau=\tau_\text{c}}
\end{align}
Substituting this expression into \eqref{uw} and equating it with \eqref{uu}, yields
\begin{align}\label{EE}
E'=-\frac{H}{R_\text{c}(\dot{H}+H^2)}\big|_{\tau=\tau_\text{c}}=\frac{-2}{1+r^{2}\omega'(r)}\big|_{r=r_+}
\end{align} 
where in the second equality we have used  \eqref{den}, \eqref{p},  \eqref{cr} and Friedman's equations. Now, the Fodor's SG \eqref{ufu} is determined by the value of \eqref{EE}  as
\begin{align}
\label{kin80}
\kappa_{\mathrm{F}}^{\text{(in)}}&=-\frac{1+{r_+^2}\omega'(r_+)}{2} \frac{\dot{R}(\tau)}{R_\text{c}} 
\end{align}
showing that Fodor's SG is proportional to the stellar contraction velocity. Our results are summarized in Table \ref{table}. For two typical examples of regular BHs, Hayward and Bardeen, as well as for 4D-EGB, the external SG is shown in Table \ref{table2}.

Substituting \eqref{EE} into \eqref{KKJ},  the null vectors \eqref{pu} for $r<R$ become
\begin{align}
\label{fodelltot}
\ell^{a}& = - \frac{R}{4}\left(1+ r_+ \omega^\prime(r_+)\right) (1,1+r H(\tau),0,0)\\
n^a &=\frac{-2}{R\left(1 + r_+^2 \omega'(r_+)\right)} (1,-1 +r H(\tau),0,0)
\end{align}
and for $r>R$ 
\begin{align}
\ell^{a}& =\frac{1}{2}\left(1 + \left(r \omega(r)\right)^{-1/2}\right)(1,1- \sqrt{\frac{1}{r\omega(r)}},0,0)\\
\label{fodntot}
n^a &=\frac{1}{1 + \left(r \omega(r)\right)^{-1/2}} (1,-1-\sqrt{\frac{1}{r\omega(r)}},0,0)
\end{align}
From \eqref{feq}, it is clear that the SG \eqref{kin80} is, in general, a function of proper time
\begin{align}
\label{kapme}
\kappa_{\mathrm{F}}^{\text{(in)}}&=\frac{ 1+ r_+^2\omega'(r_+)}{2 R_\text{c} \sqrt{R(\tau) \omega(R(\tau))}}
\end{align}
It is therefore of interest to study the evolution of the SG as the star contracts. The time dependence of the radius of a star collapsing into the regular BH, \eqref{omegreg}, and 4D-EGB BH, \eqref{omegegb}, has been calculated in \cite{our1} and \cite{our2} as
\begin{align}\label{time}
\tau= \frac{2}{3}\mathcal G(R)
\end{align}
in which
\begin{align}
\mathcal G(R)_{\text{reg}}&=R_{0}^{\frac{3}{2}}{}_{2}F_{1}[-\frac{n}{2},-\frac{n}{2},1-\frac{n}{2},-\beta R_{0}^{\frac{-3}{n}}]\nonumber\\\label{greg}
&-R_{ }^{\frac{3}{2}}{}_{2}F_{1}[-\frac{n}{2},-\frac{n}{2},1-\frac{n}{2},-\beta R^{\frac{-3}{n}}]
\end{align}
\begin{align}
\mathcal G(R)_{\text{EGB}}&=\sqrt{32 \pi  \alpha}\bigg((1+\frac{64\pi{\alpha}}{{R}^3_{0}})^{1/2}-1\bigg)^{-1/2}\nonumber\\
&-\sqrt{32 \pi  \alpha}\bigg((1+\frac{64\pi{\alpha}}{{R}^3})^{1/2}-1\bigg)^{-1/2}\nonumber \\
+& \sqrt{16 \pi  \alpha}\Big(\tan^{-1}[\frac{1}{\sqrt{2}}\big((1+\frac{64\pi{\alpha}}{{R}^3})^{1/2}-1\big)^{1/2}]
\notag\\
&-\tan^{-1}[\frac{1}{\sqrt{2}}\big((1+\frac{64\pi{\alpha}}{{R}^3_0})^{1/2}-1\big)^{1/2}]\Big)\label{gegb}
\end{align}
where  ${}_2 F_1$ is the hypergeometric function and the integration constant $R_{0}$ is chosen such that $\tau=0$ at $R=R_{0}$.
Combining the above expressions with \eqref{omegreg}, \eqref{omegegb}, \eqref{rhreg} and \eqref{rhegb}, we thus find 
the time dependence of the SG. It is worth noting that according to \eqref{time}-\eqref{gegb}, in the extremal case (i.e. $\beta=4/27$ and $\alpha=1/64\pi$) the radius of the star reaches the extremal horizon in a finite time, and the outer SG \eqref{uw} becomes zero in this case.

\begin{figure}[h]
\centering
\includegraphics[width=3.3in]{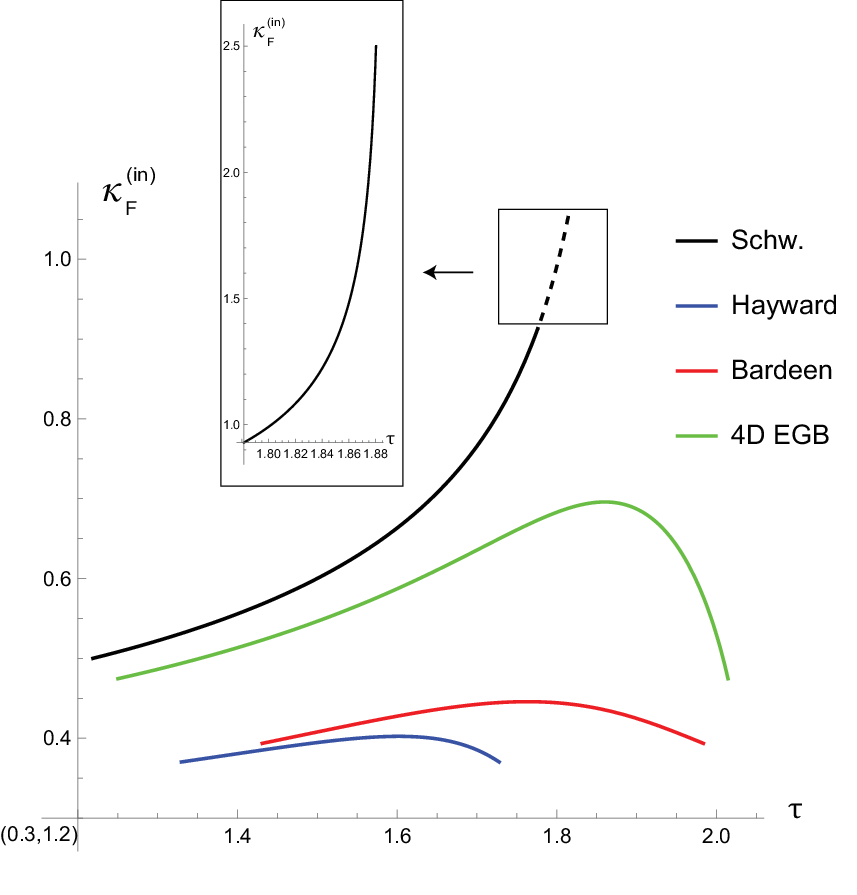}
\caption{Evolution of the SG as a function of proper time for Schwarzchild $(\omega = 1)$, Hayward $(n=1)$, Bardeen $(n=3/2)$ and 4D-EGB BHs. 
We set $\beta = 0.09$ for the regular BH and $\alpha = 0.0005$ for the 4D-EGB BH. For the Schwarzschild BH, the SG grows uniformly and diverges at the singularity, where the TH disappears.
For the regular and 4D-EGB BHs, however, the SG is finite.
Its values at $r_-$ and {$r_+$} are the same. In between, it has a maximum value.
The variables are dimensionless with respect to the Schwarzschild radius.}
\label{fig3}
\end{figure}

The evolution of the 
SG is shown in figure \ref{fig3} in the time interval $[\tau_c,\tau_e]$. The TH appears at $\tau_c$ and  then vanishes at $\tau_e$.  The maximum of the SG occurs when the surface acceleration of the star is zero. It should be remembered that in static spacetime, SG is actually the force that an observer at infinity would have to exert to keep an object at the horizon. This description does not hold in the dynamical case. Here we see that, according to \eqref{kin80}, the SG of a spherically symmetric collapsing star is maximized when its radial acceleration is zero. 
This happens on a surface between $r_ +$ and the TH of the star.
In other words, inside the star, according to Friedman's equation
\begin{equation}\label{nnew}
\ddot{R}=-\frac{4\pi}{3}(\rho+3P)
\end{equation}
the SG is a function of the density and pressure of the star. A comparison of \eqref{kin80} and \eqref{nnew} shows that the surface gravity is at a maximum when the SEC is marginally valid. This is somewhere between $r_+$ and TH of the star. Moreover, in the 4D EGB and regular BHs, the SG of the TH increases when the SEC is satisfied, and vice versa. This is in contrast to the case of gravitational collapse into the Schwarzschild BH, where the SG of the TH always increases. This is because the SEC is not violated at any radius of the star.

For a star collapsing to a near extremal BH, the time interval in figure \ref{fig3} becomes very small. In the limit of an extremal BH, the SG exists only at a certain value of the proper time and its value is zero. This can be well deduced from \eqref{kin80} where the numerator is zero for an extremal BH. 
\section{Extremality condition}
\label{sec8}
The extremality condition for stationary BHs is that the Killing SG must be zero. This condition is equivalent to the fact that in \eqref{MMBB} $f'(r)=f(r)-1=0$ has a double root $r_+=r_-=r_{\text{ex}}$ and there is no trapped surface. A coordinate-invariant definition of extremality for dynamical horizon, given in \cite{Pielahn:2011ra} as
\begin{align}\label{vvv}
n^{a}\nabla_{a}\theta_{\ell}=0.
\end{align}
By comparing \eqref{MMBB} and \eqref{viss}, the outgoing  radial null vector field of Nielsen-Visser \eqref{gggg} can be written as
\begin{align}\label{pou}
\ell^a&=(1,1-f(r,\tau),0,0)\\
n^a&=(1,-1-f(r,\tau),0,0)
\end{align}
Then performing a simple calculation leads to
\begin{align}
\theta_{\ell}&=\nabla_{a}\ell^{a}-\kappa=\frac{2}{r}(1-f(r,\tau))\\
n^a \nabla_a \theta_\ell&=-\frac{2\dot{f}}{r}-2(1+f(r,\tau))[\frac{-1}{r^2}+\frac{f(r,\tau)}{r^2}-\frac{f'(r,\tau)}{r}]
\end{align}
where $f=1$ at the horizon.
From \eqref{eq:LL}, the extremality condition \eqref{vvv} then reduces to
\begin{align}
0=n^a \nabla_a \theta_\ell\big|_{r_{\text{ex}}}=\frac{-2}{r_{\text{ex}}}(r_{\text{ex}}\omega'(r_{\text{ex}})+\omega(r_{\text{ex}}))
\end{align}
for the outer BH metric. This is exactly the same as \eqref{poi}. Recall that from \eqref{MMBB}, $f'(r)=0$ for an extremal static BH,  so  the extremality condition is trivially satisfied and this is equivalent to setting the SG \eqref{poi} of the outer horizon to zero.
For the dynamical metric inside the star 
\begin{align}
0=n^a \nabla_a \theta_\ell\big|_{r_{\text{TH}}}=2(\dot{H}(\tau)+2H^2(\tau))
\end{align}
Therefore the vanishing of this value is not equivalent to the vanishing of the SG of the TH (\ref{asd}).

Performing  a similar calculations in Fodor's approach using (\ref{pu}), we find that
\begin{align}\label{KK}
0&=n^a \nabla_a \theta_\ell\big|_{r_{\text{ex}}}=\frac{2D}{r}(\dot{C}(1-f)-C\dot{f})-2 D(1+f(r,\tau))\nonumber\\
&[\frac{C'}{r}-\frac{C'f(r,\tau)}{r}-\frac{C}{r^2}+\frac{C f(r,\tau)}{r^2}-\frac{C f'(r,\tau)}{r}]\bigg|_{r_{\text{ex}}}\nonumber\\
&=\frac{-4 D(r_{\text{ex}},\tau)}{r_{\text{ex}}} \kappa_{\mathrm{F}}^{\text{(out)}}.
\end{align}
Thus, for the $r_{\text{ex}}$ and TH of the star, we have
\begin{align}\label{KK1}
0&=n^a \nabla_a \theta_\ell\big|_{r_{\text{ex}}}=-2 D(1+f(r,\tau))\nonumber\\&[\frac{C'}{r}-\frac{C'f(r,\tau)}{r}-\frac{C}{r^2}+\frac{C f(r,\tau)}{r^2}-\frac{C f'(r,\tau)}{r}]\bigg|_{r_{\text{ex}}}
\nonumber\\&=\frac{-4 D(r_{\text{ex}},\tau)}{r_{\text{ex}}} \kappa_{\mathrm{F}}^{\text{(out)}}.\\
0&=n^a \nabla_a \theta_\ell\big|_{r_{\text{TH}}}=2 D(r_{\text{TH}},\tau)C(r_{\text{TH}},\tau)(\dot{H}+2H^2)\label{KKK}
\end{align}
where the last equality of \eqref{KK1} is a result of \eqref{pp} and gives $\kappa_{\mathrm{F}}^{\text{(out)}}=0$. However, 
a comparison of (\ref{KKK}) with \eqref{kin80} shows that the inner SG is not zero if the extremality condition is satisfied inside the star.
\section{The first law of thermodynamics}\label{sec9}
In this section, we will consider the first law of thermodynamics for the  evolving TH. 
Using the time derivative of the Misner-Sharp mass, the first law of thermodynamics takes the form
\begin{align}\label{law}
\frac{\kappa}{8\pi}\frac{dA_{\text{TH}}}{d\tau}=\frac{dM_{\text{TH}}}{d\tau}-w_{0}\frac{dV_{\text{TH}}}{d\tau}
\end{align}
where  $-w_{0}\frac{dV_{\text{TH}}}{d\tau}$ is the work term and the function $w_{0}$ must be determined. For each slice of constant time, the Misner-Sharp mass within the TH is
\begin{align}
M_{\text{TH}}=\frac{4}{3}\pi \rho r_{\text{TH}}^3=\frac{r_{\text{TH}}^{3}H^2}{2}
\end{align}
with derivative
\begin{align}\label{M}
\frac{dM_{\text{TH}}}{d\tau}=\frac{3}{2}r_{\text{TH}}^{2}\dot{r}_{\text{TH}}H^2+H\dot{H}r_{\text{TH}}^{3}=\frac{\dot{H}}{2 H^2}
\end{align}
Moreover
\begin{align}\label{A}
\frac{dA_{\text{TH}}}{d\tau}&=8\pi r_{\text{TH}}\frac{dr_{\text{TH}}}{d\tau}=-\frac{8\pi \dot{H}}{H^3}\\
\frac{dV_{\text{TH}}}{d\tau}&=4\pi r_{\text{TH}}^{2}\frac{dr_{\text{TH}}}{d\tau}=\frac{4\pi \dot{H}}{H^4}
\end{align}
As mentioned before, in the Nielsen-Visser \cite{vissern} approach, the SG is obtained by assuming that the work term in the first law of thermodynamics does not exist for the TH. But we should be careful that
\eqref{semi f} is the first law of thermodynamics with a partial time derivative of mass and surface area which gives a
different result compared to the situation where we use the total derivative. In the latter case, substituting \eqref{M} and \eqref{A} into the first law \eqref{law} gives
\begin{align}
\kappa=\frac{-H}{2}
\end{align}
which is the SG of the dynamical horizon,
as defined in \cite{Ashtekar:2002ag} and differs from the Nielsen-Visser SG \eqref{asd}, as expected.

A similar calculation for Hayward's SG \eqref{gri}, leads to
\begin{align}
w_{0}=\frac{-(\dot{H}+3H^2)}{8\pi}=\frac{P-\rho}{2}
\end{align}
and for the SG (\ref{ufu}) derived from Fodor’s approach
\begin{align}
w_{0} & =\frac{-H}{8\pi}(H+2 \kappa_{\mathrm{F}}) \nonumber \\
& =\frac{-\rho}{3}\left(1+\frac{2\exp{{(\sqrt{\frac{8\pi}{3}}\int_{\tau_\text{c}}^{\tau} \sqrt{\rho(\tau)} d\tau)}}}{E'}\right) \nonumber \\
&=\frac{-\rho}{3}\left(1+\frac{2R(\tau)}{E' R_\text{c}}\right)
\end{align}
where in the second equality we have used the Friedmann equation and the third equality is written in terms of the star's radius.

From our discussion above, it is easy to see that different definitions of SG lead to different versions of the first law of thermodynamics, as expected.
In \cite{Hayward:1997jp}, Hayward writes the laws of thermodynamics for THs and uses the Kodama vector to define SG, resulting in a non-zero work term in the first law of thermodynamics.
Ignoring the work term,
Nielsen and Visser in \cite{vissern} found a relationship similar to the first law by partial differentiation of the Misner-Sharp mass and defining the surface change coefficient as SG.
On the other hand, the first law of thermodynamics is expressed in terms of the total derivative of the thermodynamic quantities.
Therefore, the result of \cite{vissern} may not be applicable without considering the work term in the first law of thermodynamics with total derivatives.
For example in \cite{Fodor:1996rf} using Einstein’s equations, it is shown that Fodor’s definition of SG in the first law of thermodynamics is not applicable without considering the work term.
Therefore, in the above, we have considered the work term in the first law of thermodynamics for Fodor’s definition of SG.
As mentioned above, if we take the total time derivative of the horizon equation or the Misner-Sharp mass, regardless of the work term, the coefficient of variation of the area of the horizon will be equal to $1/2r_{TH}$ , which is the SG of the dynamical horizon and defined in \cite{10JFT}.
\section{Conclusion}
\label{sec10}
This paper considers the collapse of a star into a special class of regular and 4D-EGB BHs. The interior of the star is described by the spatially flat FRW geometry. To find the SG for evolving THs, we have considered several approaches. These include those of Hayward, Nielsen-Visser and Fodor. We have studied the SG for the outer and inner geometries using the above approaches. Since the outer geometry is static, the SG at the outer event horizon is the same in all approaches. However, due to the dynamical nature of the inner geometry, different approaches lead to different SGs. Tables \ref{table} and \ref{table2} summarize the results for regular and 4D-EGB BHs.

We have also seen that different SGs are not necessarily equal to the outer SG at the moment when the surface of the star crosses the outer horizon. Following Fodor's approach, we can use null vectors with arbitrary normalization coefficients which are then determined by cross-normalization and SG uniqueness conditions at the crossing time. The crossing time is defined as the moment at which the star's surface, the event horizon and the TH coincide with each other. In this way, we obtain the time evolution of the inner SG for regular and 4D-EGB BHs. It has a maximum when the collapsing acceleration of the star becomes zero. This is completely different from the behavior of the SG in the OSD collapse to a Schwarzschild BH where the SG evolves uniformly and then diverges asymptotically as the surface of the star approaches the singularity.

Applying the extremality condition, we find that it does not necessarily lead to zero SG for the interior geometry. Here, we saw that using the PG coordinate time, a finite time is required for a star to collapse into an extremal BH.
We then obtained the first law of BH thermodynamics for the evolving TH for each of the above mentioned approaches. This gives a different first law from that of Nielsen and Visser.
\label{concluding}

\section*{Acknowledgement}
F.B. and F.S. are grateful to the Iran National Science Foundation (INSF) for supporting this research under grant number 4021095. 
F.S. is grateful to the University of Tehran for supporting this work under a grant provided by the University Research Council.


\end{document}